# Guidelines for the next 10 years of proteomics

Marc R. Wilkins[1], Ron D. Appel[2], Jennifer E. Van Eyk[3], Maxey C. M. Chung[4], Angelika Görg[5], Michael Hecker[6], Lukas A. Huber[7], Hanno Langen[8], Andrew J. Link[9], Young-Ki Paik[10], Scott D. Patterson[11], Stephen R. Pennington[12], Thierry Rabilloud[13], Richard J. Simpson[14], Walter Weiss[5] and Michael J. Dunn[12]

1 Department of Biotechnology and Biomolecular Biosciences, University of New South Wales, Sydney, Australia
2 Swiss Institute of Bioinformatics and Geneva University, Geneva, Switzerland
3 Departments of Medicine, Biological Chemistry and Biomedical Engineering, Johns Hopkins University,
Baltimore, MD, USA
4 Department of Biochemistry, National University of Singapore, Singapore
5 Department of Proteomics, Technische Universität München, Freising-Weihenstephan, Germany
6 Institut fur Mikrobiologie, Ernst-Moritz-Arndt-Universität Greifswald, Greifswald, Germany
7 Biocenter, Div. Cell Biology, Medical University of Innsbruck, Austria
8 Roche Center for Medical Genomics, Hoffmann-La Roche, Basel, Switzerland
9 Department of Microbiology and Immunology, Vanderbilt University School of Medicine, Nashville, TN, USA
10 Yonsei Proteome Research Center and Department of Biochemistry, Yonsei University, Seoul, Korea
11 Amgen, Molecular Sciences, Thousand Oaks, CA, USA
12 Proteome Research Centre, UCD Conway Institute for Biomolecular and Biomedical Research,
University College Dublin Belfield, Dublin, Ireland
13 Laboratoire d'immunochimie, DRDC/ICH, INSERM U548, CEA-Grenoble, France
14 Ludwig Institute for Cancer Research and the Walter and Eliza Hall Institute of Medical Research,
Parkville, Victoria, Australia

Correspondence: Professor Marc R. Wilkins, School of Biotechnology
and Biomolecular Sciences, University of New South
Wales, Sydney, NSW 2052 Australia
E-mail: m.wilkins@unsw.edu.au
Fax: 161-2-9385-1483

Abbreviations: ICAT, isotope coded affinity tags; MudPIT, multidimensional protein identification technology

ABSTRACT

In the last ten years, the field of proteomics has expanded at a rapid rate. A range of exciting new technology has been developed and enthusiastically applied to an enormous variety of biological questions. However, the degree of stringency required in proteomic data generation and analysis appears to have been underestimated. As a result, there are likely to be numerous published findings
that are of questionable quality, requiring further confirmation and/or validation. This manuscript outlines a number of key issues in proteomic research, including those associated with experimental design, differential display and biomarker discovery, protein identification and analytical incompleteness. In an effort to set a standard that reflects current thinking on the necessary and desirable characteristics of publishable manuscripts in the field, a minimal set of guidelines for proteomics research is then described. These guidelines will serve as a set of criteria which editors of PROTEOMICS will use for assessment of future submissions to the Journal.

INTRODUCTION

Proteomics by name is now over 10 years old [1, 2] and many of the key technological innovations that made proteomics possible are of a similar or greater age. MS-based protein identification using PMF or LC-MS/MS [3–5] is now widely
adopted, and the transition from analysing one protein at a time to analysing highly complex mixtures has been made [6, 7]. For semi-quantitative analyses and comparisons, the proteomics researcher now has access to a variety of sophisticated techniques, including mass spectrometric approaches combined with stable isotopic labelling [7–11] and 2-D gel-based approaches combined with differential staining [12] or traditional visualisation and comparison techniques. The
variety of methods for the characterisation of protein co- and post-translational modifications is ever-expanding [12, 13], as are the means of elucidating protein-protein interactions [15–17]. The field of proteomics has grown at an astonishing
rate, and shows no sign of slowing. If anything, it appears to be gaining momentum as proteomic techniques become increasingly widespread and applied to an expanding smorgasbord of biological questions. The rapid expansion of proteomics, whilst exciting, has brought with it many technical issues. In some areas, it could be argued that we began to run before we could walk. New but immature technology has been enthusiastically applied to significant biological questions, and the results have become part of the scientific literature and databases. However, the degree of stringency required in proteomic data generation and analysis appears to have been underestimated. As a result, there are likely to be numerous published findings that are of questionable quality, requiring further confirmation and/or validation. For the future, a number of key but intersecting areas of concern can be identified, which if addressed should help improve the quality of proteomic research. These are discussed briefly below.

EXPERIMENTAL DESIGN

Proteomic techniques such as 2-D PAGE and shotgun proteomics (MudPIT/ICAT/iTRAQ) are powerful means of generating analytical data [6, 7, 11]. This analytical information is used to understand the nuances of the proteome of a biological
system, and in many cases is the basis of comparison of two or more samples. Yet the technical difficulty and high cost of data production, associated with highly time-consuming data analysis, has contributed to a position where poor experimental design is common. Many experiments that use 2-D PAGE have a low number of analytical and/or biological replicates, and users of MS differential display often assume that multiple estimates of differences generated by a single ICAT/iTRAQ experiment provide a substitute for experimental replicates. The reproducibility of
proteomic techniques used, as assayed by regression analysis, co-efficient of variation or other variance estimation techniques [18], is typically not reported. Power analyses, which can be used to infer the number of samples that should be analysed to discover a statistically significant result [18–20], are rarely undertaken. Weak experimental designs, particularly in a field where technical challenges remain in the production of high quality data, can make it difficult or impossible to determine if differences reported between two or more samples are likely to reflect variation in a biological system or are solely analytically derived.

DIFFERENTIAL DISPLAY AND BIOMARKER DISCOVERY

Proteomic techniques are frequently used for the discovery of differentially expressed proteins, including biomarkers.

These techniques can be used in a hypothesis-independent manner, making them attractive for this purpose. Whilst statistical tests are being increasingly applied to protein expression data, proteins are frequently published as differentially expressed on the basis of a two-fold or greater expression difference. Such conclusions ignore the analytical and biological variation inherent to any laboratory and the samples under study. It is also not infrequent to see proteins described as differentially expressed from use of univariate statistical tests (e.g. Student's t-test), but where the normal distribution of the data has been assumed but not tested.

This is of great concern as proteomic expression data is typically not normally distributed but is skewed, and requires transformation before many statistical tests can be applied [18, 21]. After appropriate statistical analysis, it may come to pass that a two-fold expression difference is shown to be significant for a particular protein. However, it is only through the detailed analysis of expression data, involving data normalisation, appropriate transformation, determination of the inherent variance and the use of suitable uni- and multivariate statistical tests, that this can be resolved.

PROTEIN IDENTIFICATION

The issue of erroneous protein identification is widely appreciated, and has been the subject of detailed discussion elsewhere [22]. Fortunately, the identification of individual proteins by PMF or LC-MS/MS is increasingly well supported by statistical scoring systems in tools [23–25], which greatly assists in the assignment of high confidence identifications.

For single or small groups of proteins, research manuscripts can also include relevant spectra or mass data, allowing for independent confirmation of identification results. By comparison, major issues remain in the identification of proteins from large-scale, automated experiments that analyse proteins via LC-MS/MS or LC-LC-MS/MS. Nonsystematic use of scoring cutoffs, that assess the number of peptides matching a protein and/or the quality of the match, can dramatically change the numbers of proteins deemed to have been identified from a complex sample [26, 27]. In some cases, protein identities are made on the basis of a match between one fragmented peptide and a database entry.

Clearly, ongoing vigilance will be required in this area, to minimise the misinterpretation of these complex data.

ANALYTICAL INCOMPLETENESS

Analytical incompleteness refers to a phenomenon where a technique used for the analysis of complex mixtures of peptides may only yield information for a fraction of relevant peptides in any single analytical run. For example, it has been observed that two replicate MudPIT analyses will produce two sets of protein identifications with ,65% overlap [28, 29]. Thirty-five percent of the proteins in the second analysis are likely to be novel compared to the first. A third replicate MudPIT analysis is likely to yield a set of identifications that has 80% overlap with those from the first two analyses, but with 20% new identifications. Because of the differences in proteins seen per run, it has been estimated that 10 to 12 MudPIT analyses may be necessary before a near-complete list of protein identities is generated from a single complex sample [28, 29]. This phenomenon has a substantial impact on use of LC-LC-MS/MS for qualitative biomarker discovery or differential display experiments, as the presence or absence of a protein in a particular run may reflect analytical incompleteness instead of true differences between samples. Accordingly, the comprehensive comparison of two or more proteomes by these approaches will require great care and high numbers of replicates [30]. It should finally be noted that analytical completeness is also inherent to the technique of 2-D PAGE, where it can arise from inconsistent sample loading and staining, leading to differences in the numbers of proteins detected in gel images [31]. However the paradigm of differential display via 2-D PAGE typically identifies proteins of interest only after statistical expression analysis, making this a less pressing issue.

MOVING TOWARDS NEW GUIDELINES

In the context of the above concerns, the authors of this article initiated a process to develop a set of minimum guidelines for the field, which could assist researchers in the execution of proteomics experiments and in the drafting of their manuscripts. It was intended to serve as a set of criteria which editors of PROTEOMICS could use for assessment of submissions, and to set a standard that reflects current thinking on the necessary and desirable characteristics of publishable manuscripts in the field. A set of 19 issues were identified from a broad-ranging discussion between the authors, who together have extensive experience in the development of proteomic technology (including bioinformatics) and its application to academic and industrial biomedical research. Each issue was drafted into a one-page working document. Reference was made to journals that publish gene expression analysis experiments, which have faced similar experimental design and data analysis issues. In July 2005 the Editors of PROTEOMICS were represented
at a workshop initiated by the journal of Molecular and Cellular Proteomics (MCP), which primarily sought to develop standards for preparing, reviewing and publishing of data from MS/MS experiments. Finally, a round-table meeting of
the authors of this document was held at the 4th HUPO conference in August 2005 to consider each issue, the outcomes of the MCP workshop, and other relevant discussion documents [22, 32]. The final set of guidelines, in the addendum to this document, is the result of this process.

Our aim has been to produce a set of guidelines that are brief and precise, that are easy to understand and not overly onerous to follow. We believe that the guidelines are selfexplanatory and do not require further discussion here. With the constant changes in technology, and the evolution of standard practices and means of data sharing in the proteomics community [33], we expect that these guidelines will
require updating from time to time. We welcome any thoughts or comments on this first set of guidelines and how they can be improved, and look forward to ongoing discussion with all proteomics researchers on this important issue.

ADDENDUM

Editorial and Author Guidelines for Publication in
PROTEOMICS

These guidelines outline issues that authors must follow when submitting a paper for publication in PROTEOMICS. Failure to follow these guidelines may be grounds for an Editorial decision to reject a manuscript without review.
Authors are also requested to take note of the different types of manuscripts that are suitable for PROTEOMICS as detailed in the 'Instructions to Authors'.

Experimental design and data analysis for 2-D PAGE and MS-based experiments
– The experimental design must be provided and must include details of the number of biological and analytical replicates. Only one biological/analytical replicate will not be acceptable. In clinical studies, it is highly desirable that a power analysis predicting the appropriate sample size for subsequent statistical analysis of the data is carried out.

– For expression analysis studies, summary statistics (mean, standard deviation) must be provided and results of statistical analysis must be shown. Reporting fold differences alone is not acceptable. Authors must report the following: methods of data normalization, transformation, missing value handling, the statistical tests used, the degrees of freedom and the statistical package or program used. Where biologically important differences in protein (gene) expression are reported, confirmatory data (e.g. from immunoassays) are desirable.

– For biomarker discovery/validation studies, the sensitivity and specificity of the biomarker(s) should be provided wherever possible. It is desirable that receiver operator characteristic (ROC) curves and areas under the curves are given.

Protein identification and characterization

– The method(s) used to generate the mass spectrometry data must be described, as should the methods used to create peak lists from raw MS or MS/MS data.

– The name and version of the program used for database searching, the values of critical search parameters (e.g. parent ion and fragment mass tolerance, cleavage rules used, allowance for number of missed cleavages) and the name and version of the database(s) searched must be provided.

– For each protein identified, measures of certainty (e.g. p-values) must be provided. For MS/MS, the number of peptides used to identify a protein must be given as well as the sequence and charge state of each peptide. For peptide mass fingerprinting, the number of peptides that match the sequence and the total percent of sequence coverage must be quoted. If extensive, the above information
should be collected as supplementary material which is available online.

– For experiments with large MS/MS data sets, estimates of the false positive rates are required (e.g. through searching randomised or reversed sequence databases). This information should be provided as supplementary material.

– Where post-translational modifications are reported, the methods used to discover the modification must be described. The modification should be mapped to amino acid(s) by fragmentation analysis, but reported as ambiguous if mapping to a single amino acid is not possible. For isobaric modifications, evidence for assigning a specific modification must be provided and the spectra included as supplementary material.

– Where protein sequence isoforms are reported, the peptide sequence that matches the unique amino acid sequence of a particular isoform must be provided. Fragmentation analysis of the appropriate peptides should be
described.

Bioinformatics

– Where a manuscript describes an academic database or software, it must be either freely accessible via the Internet, or downloadable and the access options must be provided. This also applies to commercial software or databases.

Supplementary material

– Supplementary material is encouraged. This includes protein identification results, expression data, and mass spectrometry peak lists. Note that all data must be in processed, not raw, form. This material will not be published in the printed journal but will be available online at the PROTEOMICS website (www.proteomics-journal.com).